\documentclass[twoside]{article}

\usepackage[accepted]{aistats2021_hacked}
\usepackage[round]{natbib}

\usepackage{amsmath, amsthm, amssymb, bbm, graphicx, multirow, mathrsfs, enumitem, cleveref}

\theoremstyle{plain}
\newtheorem{theorem}{Theorem}[section]
\newtheorem{lemma}[theorem]{Lemma}
\newtheorem{corollary}[theorem]{Corollary}
\newtheorem{proposition}[theorem]{Proposition}

\theoremstyle{definition}
\newtheorem{definition}[theorem]{Definition}

\theoremstyle{remark}
\newtheorem{example}[theorem]{Example}

\newcommand{\ve}{\varepsilon}

\newcommand{\mmin}{\mu_{\min}}
\newcommand{\smax}{\sigma_{\max}}

\newcommand{\bE}{\mathbb{E}}
\newcommand{\bN}{\mathbb{N}}
\newcommand{\bP}{\mathbb{P}}
\newcommand{\bR}{\mathbb{R}}
\newcommand{\bOne}{\mathbbm{1}}

\newcommand{\cM}{\mathcal{M}}

\newcommand{\cX}{\mathcal{X}}

\newcommand{\pc}{P^{\mathsf{PC}}}
\newcommand{\pb}{P^{\mathsf{PB}}}
\newcommand{\pbd}{P^{\mathsf{PBD}}}
\newcommand{\bsp}{P^{\mathsf{BSP}}}
\newcommand{\OPT}{\mathsf{OPT}}

\newcommand{\scRev}{\textsc{Rev}}
\newcommand{\scSrev}{\textsc{SRev}}
\newcommand{\scBdcrev}{\textsc{BdcRev}}
\newcommand{\scVal}{\textsc{Val}^+}
\newcommand{\cv}{\nu_w}

\newcommand{\up}{}
\newcommand{\down}{}
\newcommand{\Halmos}{}

\begin{document}

%

%

\twocolumn[

\aistatstitle{Reaping the Benefits of Bundling under High Production Costs}

\aistatsauthor{ Will Ma \And David Simchi-Levi }

\aistatsaddress{ Columbia University \And Massachusetts Institute of Technology } ]

\begin{abstract}
It is well-known that selling different goods in a single bundle can significantly increase revenue.  However, bundling is no longer profitable if the goods have high production costs.  To overcome this challenge, we introduce a new mechanism, Pure Bundling with Disposal for Cost (PBDC), where after buying the bundle, the customer is allowed to return any subset of goods for their costs.

We provide two types of guarantees on the profit of PBDC mechanisms relative to the optimum in the presence of production costs, under the assumption that customers have valuations which are additive over the items and drawn independently. We first provide a distribution-dependent guarantee which shows that PBDC earns at least $1-6c^{2/3}$ of the optimal profit, where c denotes the coefficient of variation of the welfare random variable. c approaches 0 if there are a large number of items whose individual valuations have bounded coefficients of variation, and our constants improve upon those from the classical result of Bakos and Brynjolfsson (1999) without costs.

We then provide a distribution-free guarantee which shows that either PBDC or individual sales earns at least 1/5.2 times the optimal profit, generalizing and improving the constant of 1/6 from the celebrated result of Babaioff et al. (2014). Conversely, we also provide the best-known upper bound on the performance of any partitioning mechanism (which captures both individual sales and pure bundling), of 1/1.19 times the optimal profit, improving on the previously-known upper bound of 1/1.08.

Finally, we conduct simulations under the same playing field as the extensive numerical study of Chu et al. (2011), which confirm that PBDC outperforms other simple pricing schemes overall.
\end{abstract}

\section{Introduction}

We study the monopolist pricing problem of a firm selling $n$ different items to a single random customer from the population.  For each item, the customer wants at most one copy, and has a \textit{valuation}, or maximum willingness-to-pay,
drawn from a known distribution.
The firm offers take-it-or-leave-it prices for every subset of items, and the customer chooses the subset maximizing her \textit{surplus}, assumed to equal to the \textit{sum} of her valuations for the items in the subset minus the subset's price.
Ties are broken in the firm's favor.
The objective of the firm is to maximize expected revenue, with the restriction that the empty subset must be priced at 0.

In the full generality of the problem, the firm has $2^n-1$ prices to set.  However, it is important to find profitable yet \textit{simple} pricing schemes that are explained by a small number of prices.  Two such schemes are \textit{Pure Components} (PC), where items are priced separately and the price of a subset is understood to be the sum of its constituent prices, and \textit{Pure Bundling} (PB), where the only option is to buy all of the items together at a fixed price.  A third scheme that captures both PC and PB is \textit{Mixed Bundling} (MB), which prices items individually, but offers a bundle discount if all of the items are purchased.

Simple pricing schemes such as PC, PB, and MB are ubiquitous in practice, and consequently their efficacy is a subject of great interest.
When $n=1$, the firm's only option is to sell the item individually, and the optimal individual price is the $p$ which maximizes $p(1-F(p))$, where $F$ is the CDF of the item's valuation.  
However, when $n>1$, bundling can often be better than individual pricing.
For example, suppose there are two products with IID valuations, each of which is $1$ w.p.~1/2, and $2$ w.p.~1/2.  If the items are sold individually, then the firm can earn at most $1$ from each item in expectation, for a total revenue of $2$.  On the other hand, if the items are sold in a bundle with price $3$, then the expected revenue would be $\frac{9}{4}$.

The key observation is that the valuation of the bundle is more concentrated around its mean than the valuation of the individual items, allowing the firm to set a price that is the maximum willingness-to-pay for a larger fraction of customers.
This power of PB over PC was first observed in the 
pioneering work of \cite{Sti63}, \cite{AY76}, \cite{Sch84}, and \cite{MMW89}.
However, bundling is not always more profitable than individual sales, especially once production costs are considered.
Taking the same example, suppose now that each item has an instantaneous production cost of 1.5.
Selling the items individually at price 2 each yields a profit of $(2-1.5)\cdot1/2=1/4$ per item, for a total of 1/2.
On the other hand, the optimal bundle price is 4, which sells w.p.~1/4, for a total profit of $(4-3)\cdot1/4=1/4$.
PB was no longer better here because the firm had to charge a high bundle price to cover its costs, and hence the customer needed to have high valuations for \textit{both} items in order to make a purchase.

Over the decades, a lot of work has been done to compare the profit of PB vs.\ PC.  \cite{AY76} write, ``The chief defect of Pure Bundling is its difficulty in complying with Exclusion,'' where Exclusion refers to the principle that a transfer is better off not occurring when the consumer's valuation is below the producer's cost.  It is observed in \cite{Sch84} for the case of bivariate normal valuations that PB is better when mean valuations are high compared to costs.  \cite{BB99} prove that bundling a large number of goods can extract almost 100\% of the total welfare, but this is crucially dependent on the items being ``information goods'', i.e.\ goods with no production costs.  \cite{FN06} characterize conditions under which PB outperforms PC for a fixed number of items, again highlighting the importance of low costs.

The indisputable conclusion from all this work is that high costs are the greatest impediment to the power of bundling.  However, in this paper we argue that firms can reap the benefits of bundling even under high production costs.
We propose a new pricing scheme called \textit{Pure Bundling with Disposal for Cost} (PBDC), where all of the items are sold as a bundle, but the customer is then allowed to \textit{return any subset of items} for a refund equal to their total production cost.

PBDC results in strictly higher consumer surplus than PB (with the same bundle price), because the customer can return items valued below cost for a refund equal to cost.
Meanwhile, the firm is indifferent between producing an item for its cost or returning its cost to the customer, but in the end the return option granted by PBDC entices more customers to buy the bundle, so the firm's profit is strictly increased as well.

Logistically, the extra step in PBDC of returning products and processing the refund results in additional overhead costs, which are not captured in our model.
However, the firm could enforce that the set of products to ``return'' must be decided at time of checkout, and consequently not purchased in the first place, to avoid these overhead costs.
We believe that our presentation of PBDC as a pure bundle with a return option both helps consumers choose which subset to purchase, and also helps firms analyze the profit of bundling under costs.
We now elaborate on the latter.

\subsection{Outline of Theoretical Results}

We analyze the profit of pricing according to PBDC (with an optimized bundle price) in comparison to the optimal pricing\footnote{
Technically we are comparing to the optimal randomized mechanism, which is also allowed to set prices for lotteries over items (details in Section~\ref{finite_bounds}).
} facing production costs.
If all these costs are zero, then the return option is inconsequential and PBDC coincides with the classical Pure Bundling.
In this case, our results generalize the existing results for PB, and moreover lead to \textit{improved guarantees} relative to the optimal pricing.

We also emphasize that generalizing from PB to our setting of PBDC with costs does not trivially follow from analyzing the items' valuations shifted by their costs.
This is because these shifted valuations could be negative, whereas existing analyses assume that valuations are non-negative.
Truncating negative valuations to zero could unfortunately increase the optimum against which we are comparing, since the optimal revenue is generally \textit{non-monotone} in the valuations \citep{hart2015maximal}.
Consequently, an important part of our analysis is to account for the increase in the optimum from having negative valuations.

Our first result says that the profit of PBDC is at least $1-6\cv^{2/3}$ times that of the optimum, where $\cv$ denotes the coefficient of variation of the welfare random variable.
This guarantee builds upon the results of \citet{BB99,Arm99} which say that PB extracts nearly 100\% of the optimum if there are a large number of items with independent valuations.
Indeed, in such a regime $\cv$ tends to 0 and our result says that PBDC is asymptotically optimal.
Our analysis also leads to improved constants in the convergence rate (see Section~\ref{asymptotic_bounds} for details).

Our second result says that the profit of either PBDC or PC (with optimized prices) is at least $1/5.2$ times that of the optimum.
This guarantee builds upon the line of work by \citet{hart2017approximate,LY13,BILW14} culminating in the statement that the revenue of either PB or PC is at least 1/6 times that of the optimum when there are no costs (they also show that the inclusion of PC in this statement is necessary).
We improve the bound of \citet{BILW14} from 1/6 to 1/5.2 by showing that worst cases for the \textit{core} and the \textit{tail} in their decomposition cannot occur simultaneously (see Section~\ref{finite_bounds} for details).

Our final result is an example on which neither PBDC nor PC can earn more than $\frac{3+\ln2}{3+2\ln2}\approx\frac{1}{1.19}$ times the optimum.
In fact our example contains two IID items without costs, so it applies to the classical PB setting and improves the previous-best upper bound of $\frac{12}{13}\approx\frac{1}{1.08}$ from \cite{hart2017approximate}.
We should note that a upper bound of $\frac{1}{2}$ can be found in \cite{Rub16},
but in his example PB is actually optimal if one is allowed to \textit{partition} the items into bundles,
whereas our example provides an upper bound even on the partitioning strategy (see Section~\ref{sec_upper_bound} for details).

\subsection{Summary of Numerical Experiments}

We repeat the numerical experiments from \cite{CLS08}, on the same valuation distributions and costs, with PBDC added in as a pricing scheme to be compared to PC, PB, and the \textit{Bundle-Size Pricing} (BSP) they introduce\footnote{
Prior to \cite{CLS08}, the BSP pricing scheme has appeared as ``Customized Bundling'' in \cite{HC05}.
We closely follow the experimental parameters from the working paper \cite{CLS08} but should note that the published version is \cite{CLS11}.
} which achieves over 99\% of the optimum in their simulations.
BSP is defined by parameters $\bsp_k$ which indicate the price for taking \textit{any} subset of size $k$, for all $k=1,\ldots,n$.
Note that when items have identical costs (something common in the experimental settings), PBDC is a subfamily of BSP.

Nonetheless, our experiments show that PBDC still attains between $97.5\%$ and $100\%$ of the (nearly optimal) BSP profit in these settings.
On the other hand, if costs are allowed to vary at all, then PBDC becomes the best-performing pricing scheme by far.  In fact, the worst case for PBDC is the aforementioned setting where it attains $97.5\%$; contrast this with $79.9\%$, $16.8\%$, $59.5\%$ for PC, PB, BSP respectively in their worst-case settings.  In addition to being profit-maximizing, PBDC also achieves excellent total surplus in our simulations,
and scales well with the number of items (see Section~\ref{sect_numerical_experiments} for details).

All in all, since BSP contains PBDC as a subfamily in the special case of identical costs, our theoretical guarantees for PBDC provide the first technical explanation for some of the empirical successes for BSP found by \cite{CLS08}.
On the other hand, our experiments show that it is possible to outperform BSP on the instances in \cite{CLS08} where costs are heterogeneous.
Finally, we remark that this is all despite PBDC being faster to optimize over (having 1 parameter instead of $n$) and simpler to interpret.
One could also consider an extension of PBDC where each item's return refund is optimized, instead of pegged to its production cost.
However, such an optimization problem is generally non-trivial \citep[cf.][]{li2020convex}.  We leave its solution as future work that would be a further improvement of PBDC.

\subsection{Further Related Work}
Simple mechanisms and bounds on their performance, in the special case of a single buyer, has been an active area of research over the past decade \citep{hart2017approximate,hart2013menu,hart2015maximal,LY13}.
Our distribution-free lower bound of 1/5.2 improves the celebrated 1/6-guarantee originally proved by Babaioff et al. in 2014.
There has since been many generalizations of their result, to multiple buyers \citep{Yao15}, sub-additive buyers \citep{rubinstein2018simple}, buyers with common-base-value \citep{bateni2015revenue} or proportional \citep{cai2019simple} complementary valuations, among others, which are described in \citet{BILW14}.
These papers consider setting which are more general settings than ours in some ways, but to our knowledge, none of them consider costs, or have a guarantee better than ours of 1/5.2.

Outside of mechanism design, bundling is a very broad topic whose study was pioneered by \citet{Sti63,AY76,Sch84,MMW89}.
The problem of computing optimal bundle prices is addressed in \citet{Wil93}.
Bundling can also have an impact on supply chain fulfillment, the way in which goods are marketed, for which we respectively refer to \citet{ernst1999effects,VM09} and the reference therein.


\subsection{Preliminaries}
\label{sect_setup}

A firm has $n$ different items for sale.  For each $i$, the cost incurred by the firm for selling item $i$ is $c_i\ge0$.  $c_i$ can be thought of as an instantaneous production cost, the opportunity cost of saving the inventory for someone else, or the value of the item to the seller.

Each of the firm's customers has a valuation vector $x\in\bR^n_{\ge0}$ for the items.  A customer wants at most one of each item, and her utility for a subset of items $S$ is $\sum_{i\in S}x_i$.  $x$ is a random vector drawn from a distribution $D$ representing the valuation vectors across the population.
The valuations are said to be \textit{independent} if $D$ is a product distribution $D_1\times\cdots\times D_n$, with $x_i$ drawn independently from $D_i$ for all $i=1,\ldots,n$.
The firm's objective is to maximize its expected profit.

The firm sells the items by posting for every $S\subseteq\{1,\ldots,n\}$ the price $P(S)\ge0$ that must be paid to receive exactly the subset of items in $S$, with $P(\emptyset)=0$.
A customer with valuation vector $x$ purchases a surplus-maximizing subset $S^*\in\text{argmax}_S(\sum_{i\in S}x_i-P(S))$, in which case the firm earns profit $P(S^*)-\sum_{i\in S^*}c_i$.
Ties are broken in the firm's favor.

A \textit{pricing scheme} is a restriction on the complexity of the price function $P$, by forcing it to be defined by a small number of prices in a way that is simple for the customer to understand.
We now recap some pricing schemes from the literature:
\begin{enumerate}
\item \underline{Pure Components (PC)}: the items have individual prices $\pc_1,\ldots,\pc_n\ge0$, and $P(S)=\sum_{i\in S}\pc_i$ for any subset $S$.
\item \underline{Pure Bundling (PB)}: there is a single bundle price $\pb\ge0$, and $P(S)=\pb$ for any $S\neq\emptyset$, thereby making the grand bundle the only viable purchase option.
\item \underline{Bundle-Size Pricing (BSP)}: there are prices $\bsp_1\le\ldots\le\bsp_n$ based on the number of items purchased, and $P(S)=\bsp_{|S|}$ for any $S\neq\emptyset$.
\end{enumerate}

We introduce the following pricing scheme in this paper, which takes costs into account:
\begin{enumerate}[resume]
\item
\underline{Pure Bundling with Disposal for Cost (PBDC)}: there is a price
$\pbd\ge0$ for the grand bundle, and
$P(S)=\pbd-\sum_{i\notin S}c_i$ for any
$S\neq\emptyset$, thereby
returning the costs of items $i$ not in $S$ to the customer.
\end{enumerate}

A price function $P$ which can be described in the form of one of these pricing schemes is said to \textit{fall under that pricing scheme}.
The \textit{profit of a pricing scheme} is then defined as the maximum expected profit that can be earned under the restriction that $P$ must fall under that pricing scheme.
In this paper we provide guarantees on the profit of PBDC, relative to the optimal unrestricted $P$, which hold over all instances.
We now define these benchmarks on a given instance.

\begin{definition}
A \textit{problem instance} is defined by costs $c_1,\ldots,c_n$ and a distribution $D$ from which valuation vector $x$ is drawn.  Given an instance, define:
\begin{enumerate}
\item The \textit{welfare} random variable to be $w=\sum_{i=1}^n\max\{x_i-c_i,0\}$;
\item $\mu_w$ and $\sigma_w$ to respectively denote the mean and standard deviation of $w$;
\item $\cv=\sigma_w/\mu_w$ to denote the coefficient of variation of $w$, assuming $\mu_w>0$ and $\sigma_w<\infty$;
\item $\OPT$ to denote the maximum expected profit that could be earned by any price function $P$ satisfying $P(S)\ge0$ $\forall S$ and $P(\emptyset)=0$.
\end{enumerate}
\end{definition}

In Section~\ref{asymptotic_bounds} we provide guarantees on the profit of PBDC relative to $\mu_w$, which \textit{depend on} $\cv$.  Note that $\mu_w$ is easily shown to be an upper bound on $\OPT$ \citep[see e.g.][]{BB99}.
In Section~\ref{finite_bounds} we provide \textit{distribution-free} guarantees on the profit of either PBDC or PC relative to the optimal randomized selling mechanism (which is a tighter upper bound on $\OPT$).
Note that in the special case where all costs $c_i=0$, PBDC is equivalent to PB, and welfare $w=\sum_{i=1}^nx_i$.  As a result, our guarantees for PBDC generalize existing results for PB and existing results on the profit of either PB or PC.

\section{Distribution-dependent Guarantees}\label{asymptotic_bounds}

Our main result in this section is Theorem~\ref{mr1}
and its Corollary~\ref{cr1}.
Our analysis uses Cantelli's inequality and the weighted arithmetic mean-geometric mean inequality, which are described in references \citet{Lug09} and \citet{Zha08} respectively.
All proofs are deferred to \Cref{sec:pfs1}.

\begin{theorem}\label{mr1}
For all $\ve\in[0,1]$, the expected revenue of PBDC with bundle price $\pbd=(1-\ve)\mu_w$ is at least $\frac{\ve^2-\ve^3}{\ve^2+\cv^2}\cdot\mu_w$.  In particular, if
\begin{equation}\label{benchmark_price}
\ve=\frac{2\cv^{2/3}}{3\cv^{2/3}+2},
\end{equation}
then the expected revenue of PBDC is at least
\begin{equation}\label{ugly_bound}
\frac{4\mu_w}{4+24\cv^{2/3}+45\cv^{4/3}+27\cv^2}
\end{equation}
which in turn is at least
\begin{equation}\label{cvgence_rate}
(1-6\cv^{2/3})\cdot\mu_w.
\end{equation}
\end{theorem}

Lower bound~\eqref{cvgence_rate} shows that for small $\cv$, the fraction of $\mu_w$ which PBDC is able to extract is $1-\Theta(\cv^{2/3})$, recovering an existing result from \cite{BB99} and \cite{Arm99}.
However, for large $\cv$, our tighter bound~\eqref{ugly_bound} still provides a non-trivial revenue guarantee, something not accomplished in the existing analyses.
Interestingly, from~\eqref{benchmark_price} we can see that the bundle price $\pbd$ should not be set lower than $\mu_w/3$.
This is a useful managerial reference point in situations where $\mu_w$ is known but the exact demand distribution $D$ is not \citep{chen2019distribution}, in which case the optimal value for $\pbd$ cannot be computed.

Theorem~\ref{mr1} provided a guarantee based on the coefficient of variation of the welfare random variable $w$, with the guarantee approaching 100\% as $\cv\to0$.
We now see that if there are a large number of items $n$ with \textit{independent} valuations, then indeed the law of large numbers ensures $\cv$ to be small.
\begin{corollary}\label{cr1}
Suppose that $x_1,\ldots,x_n$ are independent.
Let $\mmin$ be a uniform lower bound on the mean of $\max\{x_i-c_i,0\}$ of all $i$, and let $\smax^2$ be a uniform upper bound on the variance of $\max\{x_i-c_i,0\}$ of all $i$, with $\mmin>0$, $\smax<\infty$, and $n>(\frac{\smax}{\mmin})^2$.
Then the revenue of PBDC is at least
\begin{equation*}
(1-6(\frac{\smax}{\mmin})^{2/3}\frac{1}{\sqrt[3]{n}})\cdot\mu_w.
\end{equation*}
\end{corollary}
Taking $n\to\infty$ with $\mmin$ and $\smax$ fixed, we see that PBDC extracts the entire welfare.  Note that truncating $x_i-c_i$ from below by $0$ can only increase the mean and decrease the variance, so any bounds on the mean and variance of the untruncated $x_i-c_i$ also suffice for Corollary~\ref{cr1}.

\section{Distribution-free Guarantee} \label{finite_bounds}

Our main results in this section are Theorems~\ref{mr2} and~\ref{thm_upper_bound} analyzing the profit of PBDC relative to the optimal randomized mechanism, which we now define.
All proofs are deferred to \Cref{sec:pfs2}.

\begin{definition}
For any costs $c_1,\ldots,c_n$ and valuation distribution $D$ supported on $\cX$, define $\scRev(D)$ to be the optimal objective value of the following problem:
\begin{align*}
\max\ \bE_{x\sim D}\left[s(x)-\sum_{i=1}^nc_iq_i(x)\right] \\
\text{s.t. } \sum_{i=1}^nx_iq_i(x)-s(x) \ge \sum_{i=1}^nx_iq_i(y)-s(y) && \forall x,y\in\cX \\
\sum_{i=1}^nx_iq_i(x)-s(x) \ge 0 && \forall x\in\cX \\
q(x) \in[0,1]^n && \forall x\in\cX
\end{align*}
In the LP, $q_i(x)$ denotes the probability of a customer with valuation vector $x$ receiving item $i$, for all $x$ and $i$, while $s(x)$ denotes the expected total price paid.
The first constraints impose that a customer with valuation vector $x$ cannot receive greater surplus from lying about her valuation vector being $y$ (this is known as \textit{incentive-compatibility}).
The second constraints impose that all customers receive non-negative surplus (this is known as \textit{individual-rationality}).
\end{definition}

Any feasible mechanism is equivalent to offering a menu consisting of $q$-vectors and corresponding prices~$s$ for the customer to choose from \citep[see e.g.][]{hart2013menu}, and a revenue-optimal menu must have $s\ge0$, with equality when $q$ is the zero vector.  Consequently, $\scRev(D)$ is an upper bound on $\OPT$, which is only allowed to price the deterministic subsets $q\in\{0,1\}^n$.
We now provide a guarantee on the profit of PBDC relative to $\scRev(D)$.

\begin{theorem} \label{mr2}
On any instance with $D$ independent, the profit of either PBDC or PC is at least $\frac{1}{5.2}\cdot\scRev(D)$.
\end{theorem}

Several remarks are in order.
First, $\scRev(D)$ is a tighter upper bound on $\OPT$ than the expected welfare $\mu_w$, which could be $\infty$ when $\cv$ is unaccounted for.
Second, the inclusion of PC in the statement of Theorem~\ref{mr2} is necessary, in that there are instances on which PB alone does not achieve a constant factor, as shown in \cite{hart2017approximate}.
Theorem~\ref{mr2} is a generalization of the celebrated result of \citet{BILW14}, which says that without costs, the profit of either PB or PC is at least $\frac{1}{6}\cdot\scRev(D)$.

The first contribution of Theorem~\ref{mr2} is that it applies to settings with production costs.
Our proof first eliminates these costs by analyzing the cost-adjusted valuations $x_i-c_i$ instead.
However, the cost-adjusted valuations could be negative, which existing techniques do not handle.
Furthermore, one cannot increase the negative valuations to 0 and analyze $\max\{x_i-c_i,0\}$ instead, because increasing the valuations could increase the optimum $\scRev(D)$ against which we are comparing \citep[see][]{hart2017approximate}.
Consequently, our analysis must show how to define concepts from \cite{BILW14} such as the ``marginal mechanism'' for negative valuations.


The second contribution of Theorem~\ref{mr2} is that it improves the bound from $\frac{1}{6}$ to $\frac{1}{5.2}$.  This is obtained by analyzing the \textit{core} and the \textit{tail} in the decomposition of \citet{BILW14} together, and showing that the worst case for PBDC in the core and worst case for PC in the tail cannot simultaneously occur.

\subsection{Upper Bound on the Guarantee} \label{sec_upper_bound}

Finally, we present an upper bound to complement the lower bound presented in Theorem~\ref{mr2}.
We first show how to construct an example where \textit{Mixed Bundling} (MB), the pricing scheme of selling items individually but offering a bundle discount for purchasing all the items, performs much better than either PB or PC.

\begin{example}\label{upper_bound_example}
Consider an instance with $2$ costless items, which have IID valuations distributed as follows.  There is a point mass of size $1-\rho$ at $0$, a point mass of size $\frac{\rho}{2}$ at $2$, and the remaining $\frac{\rho}{2}$ mass distributed in an \textit{equal-revenue} fashion on $[1,2)$, i.e.\ selling individually at any price in $[1,2)$ results in the same revenue.  Formally, if $Y$ is a random variable with this distribution, then
\[\bP[Y\ge y]=\begin{cases}
1 & y=0 \\
\rho & 0<y\le1 \\
\frac{\rho}{y} & 1\le y\le2
\end{cases}
\]
where the value of $\rho$ is optimized to be $\frac{3}{3+\ln2}\approx0.81$.
\end{example}

\begin{theorem}\label{thm_upper_bound}
Consider the instance in Example~\ref{upper_bound_example}.  The best possible PC revenue is $2\rho$, attained by selling individual items at any price in $[1,2]$.  The best possible PB revenue is also $2\rho$, attained by selling the bundle at the price of $2$ or $3$.  The optimal revenue is at least $2\rho(2-\rho)$; this value can be achieved by selling individual items at the price of $2$, and the bundle at the discounted price of $3$.

Therefore, neither PC nor PB, nor any mechanism that partitions\footnote{
A \textit{partitioning mechanism} first splits the items into groups, and then sells each group as a bundle at the optimal bundle price for that group.
Partitioning mechanisms capture PC (individual selling) because this is when each item is put into its own group.
They also capture PB (pure bundling) because this is when all items are put into the same group.
} the items into bundles, can obtain more than $\frac{3+\ln2}{3+2\ln2}\cdot\scRev(D)$ which is approximately
\begin{equation*}
\frac{1}{1.19}\cdot\scRev(D).
\end{equation*}
\end{theorem}

Theorem~\ref{thm_upper_bound} provides the best-known upper bound on the broad class of partitioning mechanisms, which captures both PC and PB.  There is a tighter upper bound on the performance of just PC and PB, of 1/2, due to \citet{Rub16}.
However, in his example, PC and PB both perform poorly because they do not split the items into those which are best sold individually and those which are best sold together; in his example partitioning mechanisms are optimal.
By contrast, in our example even partitioning mechanisms are suboptimal, because they do not ``price-discriminate'', i.e.\ allow customers who highly value an item to buy it for its individual price, but still give customers with lower valuations a chance of buying it as part of a discounted bundle.

\section{Numerical Experiments}
\label{sect_numerical_experiments}

In this section we repeat the numerical experiments from \cite{CLS08}, with PBDC being added as a pricing scheme to be compared.

\subsection{Procedure}

For consistency, we follow the setup from \cite{CLS08} as closely as possible.  We use the same five families of valuation distributions commonly used to model demand---Exponential, Logit, Lognormal, Normal, and Uniform.  We also use the same ranges of parameters for these families, as outlined in Table~\ref{ranges_of_parameters}.  The parameters were calibrated so that valuations across different families have similar means on average, and the highest means are around $10$ times the lowest means.  We allow for free disposal, just like \cite{CLS08}---all negative valuations are converted to $0$.  We assume that valuations are independent across items.

\begin{table}[h]
\caption{Ranges of Parameters \citep[from][]{CLS08}} \label{ranges_of_parameters}
\begin{center}
\begin{tabular}{|c|p{5.5cm}|}
\hline
\up\down Exponential & Marginal distributions are Exponential, with means chosen uniformly from $[0.2,2]$.  Thus the rates $\lambda$ are in $[0.5,5]$. \\
\hline
\up\down Logit & Marginal distributions are Gumbel, with fixed scale $\sigma=0.25$ and means chosen uniformly from $[0,2.5]$.  Thus the locations $\mu$ are in $[-0.25\gamma,2.5-0.25\gamma]\approx[-0.14,2.36]$. \\
\hline
\up\down Lognormal & Marginal distributions are Lognormal.  Logarithms of valuations are Normally distributed with means chosen uniformly from $[-1.5,1]$ and fixed variance $\sigma^2=0.25$.  Thus the original valuations have means in $[e^{-1.5+0.125},e^{1+0.125}]\approx[0.25,3.08]$. \\
\hline
\up\down Normal & Marginal distributions are Normal with means chosen uniformly from $[-1,2.5]$ and variances chosen uniformly from $[0.25,1.75]$. \\
\hline
\up\down Uniform & Marginal distributions are Uniform on $[0,b]$, where b is chosen uniformly from $[0.4,4]$.  Thus the means are in $[0.2,2]$. \\
\hline
\end{tabular}
\end{center}
\end{table}

As far as costs, we consider three scenarios:
\begin{enumerate}
\item \textit{Heterogeneous Items}: in this scenario we allow valuation distributions to fluctuate in accordance to Table~\ref{ranges_of_parameters} while costs are kept low.  Specifically, the cost of each item is set to $0.2$, except in the case of Uniform distributions, where it is set to half the item's mean valuation.  These are the same numbers used in \cite{CLS08}.
\item \textit{Heterogeneous Costs}: in this scenario we keep the valuation distributions identical while allowing costs to fluctuate.
In the cases of the Exponential, Logit, Lognormal, or Normal distributions, the valuation is fixed to have mean 1.25, 1.5, $e^{0.5+0.125}\approx1.87$, or 1.5 (with fixed variance 1) respectively; the costs are chosen uniformly from $[0,2.5]$, approximately the same range as the means.
In the case of the Uniform distribution, the costs are chosen uniformly from $0$ to $0.75$ times the maximum valuation $b$ drawn according to Table~\ref{ranges_of_parameters}.
Generally in this scenario we have chosen the fixed means to lie in the middles of the ranges from Table~\ref{ranges_of_parameters}.
\item \textit{Heterogeneous Items and Costs}: in this scenario we allow both valuation distributions and costs to fluctuate (independently) as described in the preceding scenarios.
\end{enumerate}

We compare four pricing schemes---PC, PB, BSP, and PBDC and consider $n$ from $2$ up to $6$, which captures the range of experiments in \cite{CLS08}.
For each combination of the $3$ cost scenarios, $5$ demand distributions, and $5$ options for $n$, we randomly generate $200$ instances, resulting in $15000$ total instances.  \cite{CLS08} were able to discretize the parameter space for each combination and generate $220$ instances in a grid.  While generating instances in a grid is more reliable, we have too many combinations to do so, since we allow costs to vary independently.
Our randomized approach also has the advantage of not depending on the exact grid of parameters chosen.

\subsection{Observations}

First we report the performance of the pricing schemes separated by scenario.  For each instance (out of the $15000$), we compute which of PC, PB, BSP, PBDC earns the maximum profit on that instance, and record the performance of every pricing scheme as a \textit{fraction} of this maximum.  For each scenario (out of the $3$), we report the median performance as well as $10$'th percentile performance of every pricing scheme across the $1000$ instances of each distribution family ($200$ for each of $n=2,\ldots,6$), in Table~\ref{all_scenarios}.

\begin{table}[h]\footnotesize
\caption{Median and 10'th Percentile Performancess} \label{all_scenarios}
\begin{center}
\begin{tabular}{|c|c|cccc|}
\hline
\up\down \multicolumn{2}{|c|}{Heterogeneous Items} & PC & PB & BSP & PBDC \\
\hline
\up\down \multirow{2}{*}{Exponential}
& 0.1 \%ile & .766 & .940 & \textbf{1} & .994 \\
& 0.5 \%ile & .835 & .972 & \textbf{1} & .999 \\
\hline
\up\down \multirow{2}{*}{Logit}
& 0.1 \%ile & .826 & .937 & \textbf{1} & .988 \\
& 0.5 \%ile & .873 & .992 & \textbf{1} & .998 \\
\hline
\up\down \multirow{2}{*}{Lognormal}
& 0.1 \%ile & .734 & .982 & \textbf{1} & .998 \\
& 0.5 \%ile & \textit{.799} & .996 & \textbf{1} & 1 \\
\hline
\up\down \multirow{2}{*}{Normal}
& 0.1 \%ile & .825 & .745 & \textbf{1} & .957 \\
& 0.5 \%ile & .890 & .880 & \textbf{1} & \textit{.975} \\
\hline
\up\down \multirow{2}{*}{Uniform}
& 0.1 \%ile & .904 & .834 & .940 & \textbf{.949} \\
& 0.5 \%ile & .959 & .867 & .975 & \textbf{.998} \\
\hline
\up\down \multicolumn{2}{|c|}{Heterogeneous Costs} & PC & PB & BSP & PBDC \\
\hline
\up\down \multirow{2}{*}{Exponential}
& 0.1 \%ile & .850 & .269 & .807 & \textbf{.995} \\
& 0.5 \%ile & .931 & .489 & .907 & \textbf{1} \\
\hline
\up\down \multirow{2}{*}{Logit}
& 0.1 \%ile & .815 & .063 & .245 & \textbf{.996} \\
& 0.5 \%ile & .891 & .481 & \textit{.595} & \textbf{1} \\
\hline
\up\down \multirow{2}{*}{Lognormal}
& 0.1 \%ile & .775 & .513 & .760 & \textbf{1} \\
& 0.5 \%ile & .861 & .730 & .880 & \textbf{1} \\
\hline
\up\down \multirow{2}{*}{Normal}
& 0.1 \%ile & .858 & .297 & .779 & \textbf{.982} \\
& 0.5 \%ile & .926 & .547 & .912 & \textbf{1} \\
\hline
\up\down \multirow{2}{*}{Uniform}
& 0.1 \%ile & .872 & .348 & .875 & \textbf{.948} \\
& 0.5 \%ile & .933 & .578 & .974 & \textbf{1} \\
\hline
\up\down \multicolumn{2}{|c|}{Both Heterogeneous} & PC & PB & BSP & PBDC \\
\hline
\up\down \multirow{2}{*}{Exponential}
& 0.1 \%ile & .884 & .137 & .759 & \textbf{.978} \\
& 0.5 \%ile & .964 & .403 & .926 & \textbf{1} \\
\hline
\up\down \multirow{2}{*}{Logit}
& 0.1 \%ile & .852 & .001 & .385 & \textbf{.987} \\
& 0.5 \%ile & .938 & \textit{.168} & .894 & \textbf{1} \\
\hline
\up\down \multirow{2}{*}{Lognormal}
& 0.1 \%ile & .852 & .015 & .327 & \textbf{.931} \\
& 0.5 \%ile & .970 & .245 & .887 & \textbf{1} \\
\hline
\up\down \multirow{2}{*}{Normal}
& 0.1 \%ile & .904 & .010 & .699 & \textbf{.974} \\
& 0.5 \%ile & .978 & .198 & .933 & \textbf{1} \\
\hline
\up\down \multirow{2}{*}{Uniform}
& 0.1 \%ile & .914 & .380 & .605 & \textbf{.937} \\
& 0.5 \%ile & .982 & .638 & .875 & \textbf{1} \\
\hline
\end{tabular}
\end{center}
\end{table}
\normalsize

We know from \cite{CLS08} that BSP is within $1\%$ of the optimal deterministic pricing in most of their settings, so there is minimal room for improvement under scenario~1.  In fact, PBDC is a special case of BSP when all costs are identical, and very similar to PB when costs are low.  However, as one can see in Table~\ref{all_scenarios}, PBDC still extracts close to $100\%$ of the near-optimal BSP profit under this scenario.  For Uniform valuations, PBDC is no longer a special case of BSP, since costs vary proportionally with means.  PBDC actually outperforms BSP in this setting---indeed, this is by far the worst setting for BSP listed in \citet[tbl.~5]{CLS08}, where it only extracts $91\%$ of the optimum.

Scenario~2, in which valuation distributions are identical but costs are allowed to fluctuate, really exhibits the power of PBDC, which allows customers to consume only the items they value above cost via self-selection.  PC loses out on not bundling similar items that differ only in cost, while BSP is forced to compromise between charging cheap prices which cause overinclusion loss in the high-cost items, or charging expensive prices which cause deadweight loss in the low-cost items.\footnote{
The welfare (which is fixed) can be decomposed as the sum of the firm's profit (``producer surplus''), the customer's utility minus cost (``consumer surplus''), the sum of $x_i-c_i$ over items not sold which should have been sold because $x_i>c_i$ (this is the ``deadweight loss''), and the sum of $c_i-x_i$ over items sold which should not have been because $c_i>x_i$ (this is the ``overinclusion loss'').
In general, to increase profit and consumer surplus one aims to minimize deadweight loss and overinclusion loss.
\label{econ_figures_def}
}  In \Cref{video_game_example}, we show an instance that exemplifies why BSP performs poorly when the costs in the setup from \cite{CLS08} are increased.

When both valuation distributions and costs are allowed to vary under scenario~3, PBDC is still the best strategy by a significant margin.  However, the benefits of bundling have decreased when items can be drastically different, and consequently PC has gained ground.  It seems intuitive to hypothesize that the performance of PC is inflated by the small values of $n$ we are using.  In the next subsection, we organize our reports separated by $n$, under scenario~3 where both valuation distributions and costs fluctuate.

\subsection{Separation by $n$ and Breakdown of Welfare}

In this subsection, we allow both valuation distributions and costs to fluctuate, and report averages across demand distributions, separated by $n$.  Since the distribution families we're amalgamating were calibrated to have similar means over their ranges of parameters once $n$ is fixed, it makes sense in this subsection to report average absolute profits, instead of median fractions.  We also report the ``economics figures'' defined in Footnote~\ref{econ_figures_def}, in the spirit of \cite{CLS08}, in Table~\ref{economics_table}.
The main conclusions are then summarized in graphs.

\begin{table}[h]\footnotesize
\caption{Report of Economics Figures} \label{economics_table}
\begin{center}
\begin{tabular}{|c|c|cccc|}
\hline
\up\down $n$ & Statistic & PC & PB & BSP & PBDC \\
\hline
\up &Producer Surplus&0.427&0.301&0.412&\textbf{0.432}\\
&Consumer Surplus&0.287&0.194&0.250&0.292\\
2&Total Surplus&0.714&0.495&0.662&0.724\\
&Deadweight Loss&0.192&0.351&0.224&0.183\\
\down &Overinclusion Loss&-&0.061&0.021&-\\
\hline
\up &Producer Surplus&0.655&0.395&0.630&\textbf{0.683}\\
&Consumer Surplus&0.437&0.254&0.382&0.436\\
3&Total Surplus&1.092&0.649&1.011&1.119\\
&Deadweight Loss&0.291&0.604&0.352&0.264\\
\down &Overinclusion Loss&-&0.130&0.020&-\\
\hline
\up &Producer Surplus&0.870&0.457&0.827&\textbf{0.929}\\
&Consumer Surplus&0.587&0.293&0.497&0.582\\
4&Total Surplus&1.456&0.749&1.324&1.511\\
&Deadweight Loss&0.396&0.905&0.498&0.342\\
\down &Overinclusion Loss&-&0.198&0.031&-\\
\hline
\up &Producer Surplus&1.070&0.504&1.030&\textbf{1.167}\\
&Consumer Surplus&0.705&0.297&0.595&0.703\\
5&Total Surplus&1.775&0.802&1.625&1.870\\
&Deadweight Loss&0.488&1.158&0.600&0.394\\
\down &Overinclusion Loss&-&0.304&0.039&-\\
\hline
\up &Producer Surplus&1.265&0.553&1.206&\textbf{1.409}\\
&Consumer Surplus&0.844&0.346&0.697&0.828\\
6&Total Surplus&2.108&0.899&1.902&2.237\\
&Deadweight Loss&0.587&1.440&0.736&0.459\\
\down &Overinclusion Loss&-&0.356&0.057&-\\
\hline
\end{tabular}
\end{center}
\end{table}
\normalsize

\begin{figure}[h]
\centerline{\includegraphics[width=.9\columnwidth]{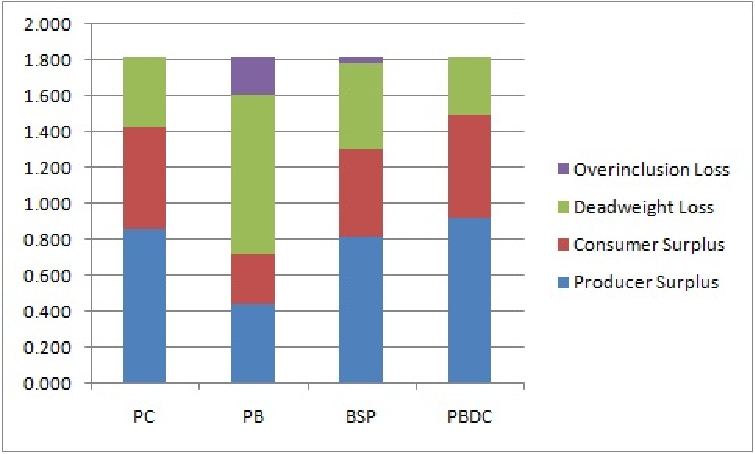}}
\caption{Breakdowns of Welfare, averaged over $n$}
\label{breakdown_of_welfare_graph}
\end{figure}

The first graph (Figure~\ref{breakdown_of_welfare_graph}) shows that although PBDC optimizes from the perspective of a selfish monopolist interested only in profit, it has a similar advantage in terms of sum of producer and consumer surplus.  Indeed, there is zero overinclusion loss, and the monopolist is encouraged to choose a bundle price low enough to accommodate most customers.  PC also incurs no overinclusion loss, but incurs more deadweight loss because it does not bundle.  PB incurs significantly more overinclusion loss than any other strategy, forcing the customer into buying every item at once.  All in all, PBDC is equally attractive from the long-term perspective of maximizing the customers' surplus.

\begin{figure}[t]
\begin{center}
\includegraphics[width=.8\columnwidth]{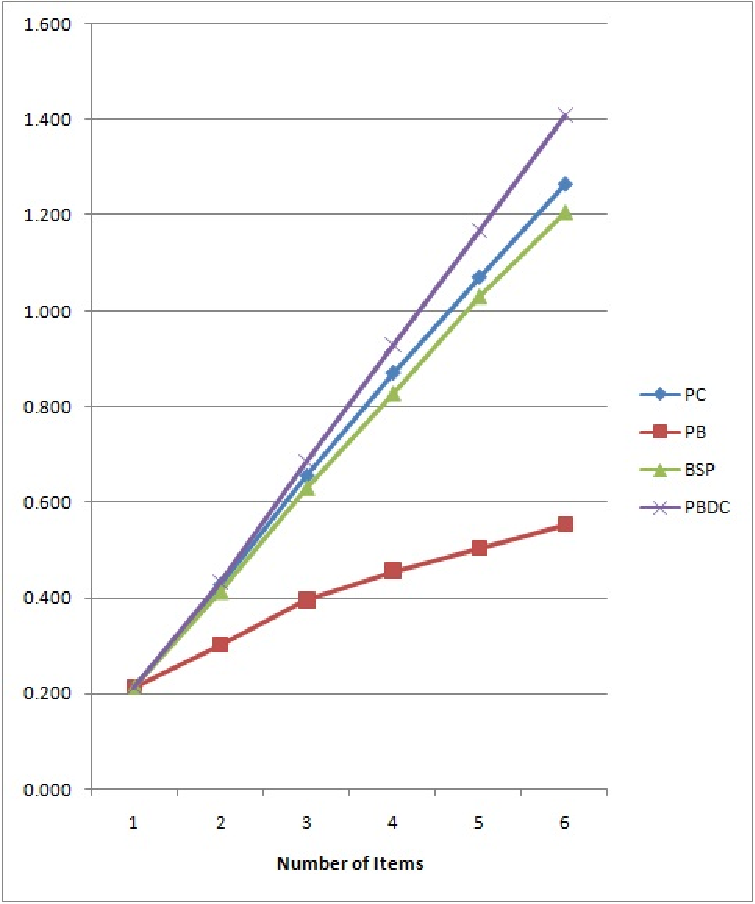}
\caption{Average Profits, as a function of $n$} \label{increase_with_n_graph}
\end{center}
\end{figure}

The second graph (Figure~\ref{increase_with_n_graph}) shows the profits of each pricing scheme as $n$ increases.  PC profits increase linearly with $n$, since items are sold separately.  Both PB and BSP profits are concave in $n$---that is, the marginal gain from having one more item to sell is decreasing.  Indeed, PB is burdened with adding to its grand bundle another item that could be valued below cost, while BSP is burdened with an additional distinct item to consider in its item-symmetric price structure.  PBDC is the only pricing scheme where the profit curve is (slightly) convex in $n$, since each item creates additional incentive for the customer to purchase the bundle, and makes the customer's total utility from purchasing more concentrated about its mean.  This confirms the hypothesis that while Table~\ref{all_scenarios} reports a small gap between PC and PBDC under scenario~3, this gap quickly widens as $n$ increases.

To summarize our numerical experiments, we considered both scenarios with low costs and scenarios with high costs, and reported median performances over $n=2,\ldots,6$ for different demand distributions.  When costs are low, PC can earn as little as $79.9\%$ of the profit of the best mechanism among PC, PB, BSP, and PBDC.
When costs are high, PB can earn as little as $16.8\%$ of the profit of the best mechanism, BSP can earn as little as $59.5\%$, and PC also falls behind as $n$ increases.
PBDC has the highest percentages overall, and is by far the most robust over different cost scenarios, always obtaining at least $97.5\%$ of the profit of the best mechanism among PC, PB, BSP, and PBDC.  We should point out that throughout our simulations, PBDC was also computationally much faster than BSP, requiring an optimization over $1$ parameter instead of $n$ parameters.

\subsubsection*{Acknowledgements}

The authors would like to thank anonymous reviewers as well as Aviad Rubinstein whose excellent suggestions helped improve the paper.
This is the full supplement to a version from the proceedings of the 24th International Conference on Artificial Intelligence and Statistics (AISTATS), 2021.

\clearpage

\bibliographystyle{abbrvnat} 
\bibliography{bibliography} 

\clearpage
\onecolumn
\aistatstitle{Reaping the Benefits of Bundling under High Production Costs: \\
Supplementary Materials}

\clearpage
\section{Missing Proofs from Section~\ref{asymptotic_bounds}} \label{sec:pfs1}

\begin{proof}[Proof of Theorem~\ref{mr1}.]
We would like to bound from above the probability that $w<(1-\ve)\mu_w$, based on the variance in $w$.
We use Cantelli's inequality instead of the more standard Chebyshev's inequality because we are only considering a one-sided tail.
We now precisely state Cantelli's inequality, reproduced from \citet{Lug09}.
\begin{lemma}\label{cantelli}
(Cantelli's Inequality) Let $X$ be a random variable with finite mean $\mu$ and variance $\sigma^2$.  Let $t$ be an arbitrary non-negative real number.  Then
\begin{equation*}
\bP[X-\mu\le-t]\le\frac{\sigma^2}{\sigma^2+t^2}.
\end{equation*}
\end{lemma}
Applying Cantelli's inequality with $t=\ve\mu_w$, we obtain that $\Pr[w<(1-\ve)\mu_w]\le\frac{\sigma^2}{\sigma^2+\ve^2\mu^2}$.  Therefore, the expected revenue is at least 
\begin{equation*}
(1-\ve)\mu\cdot(1-\frac{\sigma^2}{\sigma^2+\ve^2\mu^2})=\mu\cdot\frac{(1-\ve)\ve^2\mu^2}{\sigma^2+\ve^2\mu^2}.
\end{equation*}
The fraction of expected welfare earned is
\begin{eqnarray}
\frac{\ve^2-\ve^3}{\ve^2+\cv^2} & \ge & \frac{\ve^2-\ve^3}{\frac{2}{3}\ve^3\cv^{-2/3}+\frac{1}{3}\cv^{4/3}+\cv^2} \label{original} \\
& \ge & \frac{\ve^2-(1+\frac{2}{3}\cv^{-2/3})\ve^3}{\frac{1}{3}\cv^{4/3}+\cv^2}, \label{derivative}
\end{eqnarray}
where inequality~\eqref{original} holds because the \textit{weighted arithmetic mean--geometric mean inequality} \citep[see e.g.][]{Zha08} says
\begin{align*}
\frac{2\ve^3\cv^{-2/3}+\cv^{4/3}}{3}\ge\left((\ve^3\cv^{-2/3})^2(\cv^{4/3})\right)^{1/3}=\ve^2,
\end{align*}
while inequality~\eqref{derivative} holds because for a fraction $\frac{a}{b}$ with $0<a\le b$, subtracting the same amount less than $b$ from both the numerator and the denominator can only decrease the fraction.

Now, if we choose $\ve=\dfrac{2\cv^{2/3}}{3\cv^{2/3}+2}$, the value obtained by setting the derivative of (\ref{derivative}) to zero, then the LHS of (\ref{original}) becomes
\begin{eqnarray*}
\frac{4\cv^{4/3}(1-\frac{2}{3})}{(3\cv^{2/3}+2)^2(\frac{1}{3}\cv^{4/3}+\cv^2)} & = & \frac{\frac{4}{3}}{(2+3\cv^{2/3})^2(\frac{1}{3}+\cv^{2/3})} \\
& = & \frac{4}{4+24\cv^{2/3}+45\cv^{4/3}+27\cv^2} \\
& = & 1-6\cv^{2/3}\left(\frac{4+\frac{15}{2}\cv^{2/3}+\frac{9}{2}\cv^{2/3}}{4+24\cv^{2/3}+45\cv^{4/3}+27\cv^2}\right) \\
& \ge & 1-6\cv^{2/3}
\end{eqnarray*}
where the inequality holds because the expression in parentheses is less than 1.  This establishes both (\ref{ugly_bound}) and (\ref{cvgence_rate}), completing the proof of Theorem~\ref{mr1}.
\end{proof}

\begin{proof}[Proof of Corollary~\ref{cr1}.]
By linearity of expectation, $\mu_w\ge n\mmin$
By independence, $\sigma_w^2\le n\smax^2$.  Therefore, $\cv$ is upper bounded by $\frac{\smax}{\mmin\sqrt{n}}$, and it is easy to see from the proof of Theorem~\ref{mr1} that all of its statements continue to hold when $\cv$ is replaced by an upper bound on $\cv$.  The condition $n>(\frac{\smax}{\mmin})^2$ ensures that $\cv<1$, and the result follows immediately from substituting $\cv\le\frac{\smax}{\mmin\sqrt{n}}$ into (\ref{cvgence_rate}).
\end{proof}

\section{Proof of 1/5.2-Guarantee from Section~\ref{finite_bounds}} \label{sec:pfs2}

The first step of proving Theorem~\ref{mr2} is to eliminate the costs by analyzing the cost-adjusted valuations $x_i-c_i$ instead.  The following lemma formalizes how to perform this transformation.

\begin{lemma}\label{transformation}
The firm's problem of maximizing expected profit with distribution $D$ and costs $c$ is equivalent to the transformed problem of maximizing expected revenue with distribution $D'$, where $D'$ is the distribution $D$ shifted downward by $c_i$ in every dimension $i$.
If the firm was restricted to the pricing scheme PBDC, then in the transformed problem, the firm is restricted to the pricing scheme which charges the same price for any non-empty subset of items.
\end{lemma}

\begin{proof}[Proof of Lemma~\ref{transformation}.]
The firm's problem optimization problem over mechanisms can be rewritten as
\begin{equation*}
\begin{array}{rrclr}
\max & \bE_{x\sim D}[s(x)-q(x)^Tc] & & & \\
s.t. & q(x)^T(x-c)-(s(x)-q(x)^Tc) & \ge & q(y)^T(x-c)-(s(y)-q(y)^Tc) & \forall x,y\in\cX \\
& q(x)^T(x-c)-(s(x)-q(x)^Tc) & \ge & 0 & \forall x\in\cX \\
& q(x) & \in & [0,1]^n & \forall x\in\cX
\end{array}
\end{equation*}
Now, define $x':=x-c$, $y':=y-c$, $q'(x):=q(x+c)$, and $s'(x):=s(x+c)-q(x+c)^Tc$.  Let $\cX':=\{x-c:x\in\cX\}$, and similarly let $D'$ be the distribution $D$ shifted $c_i$ units downward in dimension $i$ for every $i\in[n]$.  We can see that the above is equivalent to
\begin{equation*}
\begin{array}{rrclr}
\max & \bE_{x'\sim D'}[s'(x')] & & & \\
s.t. & q'(x')^Tx'-s'(x') & \ge & q'(y')^Tx'-s'(y') & \forall x',y'\in\cX' \\
& q'(x')^Tx'-s'(x') & \ge & 0 & \forall x'\in\cX' \\
& q'(x') & \in & [0,1]^n & \forall x'\in\cX
\end{array}
\end{equation*}
which is identical to the original problem, without costs, on this new distribution $D'$.

Now consider the pricing scheme PBDC.  $\cM$ is restricted to be of the form $\{(\bOne_S,\pbd_0-\bOne_{[n]\setminus S}^Tc):\emptyset\neq S\subseteq[n]\}\cup\{(0,0)\}$ where $\bOne_S\in\{0,1\}^n$ is the indicator vector for items belonging to $S$.  Hence $\cM'$ is restricted to be of the form
\begin{equation*}
\{(\bOne_S,\pbd_0-\bOne_{[n]\setminus S}^Tc-\bOne_S^Tc):\emptyset\neq S\subseteq[n]\}\cup\{(0,0)\}=\{(\bOne_S,\pbd_0-\bOne_{[n]}^Tc):\emptyset\neq S\subseteq[n]\}\cup\{(0,0)\}.
\end{equation*}
Put in words, $\cM'$ must belong to the class of menus that offer the same price for any non-empty subset of items.  
\end{proof}

In the latter part of Lemma~\ref{transformation}, the fact that the customer can choose to take a subset of items instead of taking all the items is important, because valuations $x'_i$ can be negative.
We will interpret this as being equivalent to PB under the assumption of \textit{free disposal} of items.  (In the original description of PB, the distinction of free disposal was irrelevant, since all valuations were non-negative.)

We now proceed to prove Theorem~\ref{mr2} where we will hereafter use $x_i$ to refer to the transformed, potentially negative valuations.  Furthermore, we normalize these valuations so that the revenue of PC is 1.

\subsection{The Core-Tail Decomposition}

We use the core-tail decomposition of \cite{BILW14}, with the original idea coming from \cite{LY13}.  We will cut up the domain of the joint distribution and consider the conditional distributions on the smaller subdomains.  Below, we introduce the notation for working with these distributions on smaller subdomains.  One should get comfortable with the idea that some of the distributions defined could be the null distribution, if they were distributions conditioned on a set of measure $0$, or a product over an empty set of distributions.  The product of a null distribution with any other distribution is still a null distribution.

\begin{definition}
We make the following definitions.
\begin{enumerate}
\item For all $i\in[n]$, let $r_i$ denote the optimal revenue earned by selling item $i$ individually (by our normalization, $\sum_{i=1}^nr_i=1$).
\item Let $D_i^C$ (the ``core'' of $D_i$) denote the conditional distribution of $D_i$ when it lies in the range $(-\infty,1]$.
\item Let $D_i^T$ (the ``tail'' of $D_i$) denote the conditional distribution of $D_i$ when it lies in the range $(1,\infty)$.
\item Let $p_i:=\bP_{x_i\sim D_i}[x_i>1]$, the probability item $i$ lies in its tail.
\item Let $A\subseteq[n]$ represent a subset of items, usually the items whose valuations lie in their tails.
\item Let $D_A^T:=\times_{i\in A}D_i^T$, the product distribution of only items in their tails.
\item Let $D_A^C:=\times_{i\notin A}D_i^C$, the product distribution of only items in their cores.
\item Let $D_A:=D_A^C\times D_A^T$, the conditional distribution of $D$ when exactly the subset $A$ of items lie in their tails.  Let $p_A$ be the probability this occurs, which is equal to $(\prod_{i\notin A}(1-p_i))(\prod_{i\in A}p_i)$, by independence.
\item Let $x_i^+:=\max\{x_i,0\}$.
\item For any valuation distribution $S$, let $\scVal(S):=\sum_i\bE_{x\sim S}[x_i^+]$, which is the expected welfare after the transformation from costs to negative valuations.  Note that the sum is only over the admissible $i$ if $S$ is a distribution on a smaller subdomain.
\item Let $\scRev(S)$ denote the optimal revenue obtainable from valuation distribution $S$ via any Incentive Compatible and Individually Rational mechanism, which could include lotteries.
\item Let $\scSrev(S)$ denote the optimal revenue of any pricing scheme falling under the class of separate sales (Pure Components).
\item Let $\scBdcrev(S)$ denote the optimal revenue of any pricing scheme falling under the class of PBDC.
\end{enumerate}
(It is understood that  $\scVal,\scRev,\scSrev,\scBdcrev$ are $0$ when evaluated on the null distribution.)
\end{definition}

\subsection{Lemmas for Negative Valuations}
\label{lemmas}

We need to extend the statements of lemmas from \cite{hart2017approximate}, \cite{LY13}, and \cite{BILW14} to handle negative valuations.

\begin{lemma}(Marginal Mechanism)
\label{marginal_mechanism}
Let $S,S'$ be (potentially negative) valuation distributions over disjoint sets of items.  Then
\begin{equation*}
\scRev(S\times S')\le\scVal(S)+\scRev(S')
\end{equation*}
\end{lemma}
The Marginal Mechanism tells us that when selling a group of independent items, we cannot do better than breaking off some items individually, extracting the entire welfare from those items, and selling the remaining items as a group.

\proof{Proof of Lemma~\ref{marginal_mechanism}.}
Consider the following mechanism for selling to a buyer with valuations drawn from $S'$.  First, sample a value $v\sim S$, and reveal to the buyer these make-believe valuations for the items in $S$.  Then run a mechanism obtaining $\scRev(S\times S')$ on this buyer, with the modification that whenever the buyer would have received an item $i$ from the support of $S$, instead she will receive (or pay) money equal to $v_i$.  By independence, this modified mechanism on the buyer with valuations drawn from $S'$ is IC and IR (a buyer with valuations $S'$ will choose the same menu entry under the modified mechanism as a buyer with valuations $S\times S'$ would have chosen under the original mechanism) and we will obtain $\scRev(S\times S')$, but then have to settle for the items in $S$.  The most we stand to lose in the settlement is $\sum_iv_i^+$ (each item $i$ in $S$ is transferred in full whenever $v_i\ge0$, and not transferred when $v_i<0$), so this amount is upper bounded in expectation by $\scVal(S)$.  Therefore, the optimal revenue from $S'$ is at least $\scRev(S\times S')-\scVal(S)$, completing the proof of the lemma.
\Halmos\endproof

\begin{lemma}(Subdomain Stitching)
\label{subdomain_stitching}
Let $S$ be a product distribution over valuations, with support $\cX\subseteq\bR^m$ for some $m\in\bN$.  Let $\cX_1,\ldots,\cX_k$ form a partition of $\cX$ inducing conditional distributions $S^{(1)},\ldots,S^{(k)}$, respectively, and let $s_j=\bP_{x\sim S}[x\in\cX_j]$.  Then
\begin{equation*}
\scRev(S)\le\sum_{j=1}^ks_j\scRev(S^{(j)})
\end{equation*}
\end{lemma}
Intuitively, Subdomain Stitching says that revenue can only increase if we sell to each subdomain separately, since we can use a different mechanism for each subdomain that specializes in extracting the welfare from that customer segment.

\proof{Proof of Lemma~\ref{subdomain_stitching}.}
Let $M$ be an optimal mechanism obtaining $\scRev(S)$, and for any valuation distribution $S'$, let $\scRev_M(S')$ denote the expected revenue obtained from mechanism $M$ when the buyer's valuation is drawn from $S'$.  Clearly $\scRev(S)=\sum_{j=1}^ks_j\scRev_M(S^{(j)})$, and furthermore for all $j\in[k]$, $\scRev_M(S^{(j)})\le\scRev(S^{(j)})$ since M is an IC-IR mechanism for selling to $S^{(j)}$, completing the proof of the lemma.
\Halmos\endproof

\begin{lemma}
\label{subdomain_reverse}
Let $S$ be a product distribution over valuations, with support $\cX\subseteq\bR^m$ for some $m\in\bN$.  Let $\cX'$ be a subset of $\cX$ inducing conditional distribution $S'$, and let $s'=\bP_{x\sim S}[x\in\cX']$.  Then
\begin{equation*}
\scRev(S)\ge s'\scRev(S')
\end{equation*}
\end{lemma}
While Subdomain Stitching places an upper bound on $\scRev(S)$, Lemma~\ref{subdomain_reverse} places a lower bound on $\scRev(S)$ based on the optimal revenue of any single subdomain.

\proof{Proof of Lemma~\ref{subdomain_reverse}.}
Consider an optimal mechanism for $S'$, and extend this to an IC-IR mechanism on $S$ by allowing the buyer to report a value in $\cX'$ maximizing her utility.  With probability $s'$, the buyer's valuation will actually be drawn from $S'$ and we will obtain revenue $\scRev(S')$; otherwise, we still earn a non-negative revenue, since the mechanism never admits a negative payment.  Therefore, the optimal revenue for $S$ is at least $s'\scRev(S')$, completing the proof of the lemma.
\Halmos\endproof

\begin{lemma}
\label{number_of_items}
Let $S$ be a product distribution over $m$ independent (potentially negative) valuations, for some $m\in\bN$.  Then
\begin{equation*}
\scRev(S)\le m\cdot\scSrev(S)
\end{equation*}
\end{lemma}
While selling $m$ items together can definitely be better than selling them separately, this lemma tells us it can be no more than $m$ times better.

\proof{Proof of Lemma~\ref{number_of_items}.}
We proceed by induction.  The statement is trivial when $m=1$.  Now, suppose we have proven the statement for $m$ valuations, and we will prove it for $m+1$ valuations.

Partition the support $\cX\subseteq\bR^{m+1}$ of $S$ into $\cX_1$ and $\cX_2$, where $\cX_1:=\{x\in\cX:x_1\ge\max\{x_j,0\}\ \forall\ j=2,\ldots,m+1\}$ and $\cX_2:=\cX\setminus\cX_1$.  Let $s_1$ denote the probability a value sampled from $S$ lies in $\cX_1$, and let $S_1$ be its distribution conditioned on this event.  Define $s_2,S_2$ respectively.  Subdomain stitching tells us $\scRev(S)\le s_1\scRev(S^{(1)})+s_2\scRev(S^{(2)})$.  Our goal is to separately show that $s_1\scRev(S^{(1)})\le(m+1)\scSrev(S_1)$ and $s_2\scRev(S^{(2)})\le(m+1)\scSrev(S_{-1})$.

Now, applying Marginal Mechanism on $S^{(1)}$ and multiplying both sides of the inequality by $s_1$, we get $s_1\scRev(S^{(1)})\le s_1\scVal(S^{(1)}_{-1})+s_1\scRev(S^{(1)}_1)$.  By considering a distribution that samples $v\sim S$ but only outputs $v_1$, we can use Lemma~\ref{subdomain_reverse} to show that $s_1\scRev(S^{(1)}_1)\le\scRev(S_1)$.  To bound $\scVal(S^{(1)}_{-1})$, consider the following mechanism for selling just item $1$: sample $v_{-1}\sim S_{-1}$, and set the price to be $\max_{i=2}^{m+1}\{\max\{v_i,0\}\}$.  Since the buyer's valuation is drawn from $S_1$, by independence, we get a sale with probability exactly $s_1$.  Furthermore, $\max_{i=2}^{m+1}\{\max\{v_i,0\}\}\ge\frac{1}{m}\sum_{i=2}^{m+1}\max\{v_i,0\}$, so conditioned on us getting a sale, the expected payment is at least $\frac{1}{m}\scVal(S^{(1)}_{-1})$.  We have proven $\scRev(S_1)\ge\frac{s_1}{m}\scVal(S^{(1)}_{-1})$, hence $s_1\scRev(S^{(1)})\le(m+1)\scRev(S_1)=(m+1)\scSrev(S_1)$, as required.

It remains to bound $s_2\scRev(S^{(2)})$, and using Marginal Mechanism and Lemma~\ref{subdomain_reverse} in the same way as before, we obtain that it is no more than $s_2\scVal(S^{(2)}_1)+\scRev(S_{-1})$.  Consider the following mechanism for selling items $2,\ldots,m+1$: sample $v_1\sim S_1$, and set the individual price for each item $2,\ldots,m+1$ to be $\max\{v_1,0\}$.  Note that the probability of getting at least one sale is less than $s_2$, since even when there is some $j=2,\ldots,m+1$ such that $v_1<\max\{x_j,0\}$, it is possible for both $v_1,x_j$ to be negative.  However, in this case $\max\{v_1,0\}=0$, so not getting a sale is still equivalent to getting at least one sale for $\max\{v_1,0\}$.  Therefore, we can think of it as we get at least one sale with probability $s_2$, in which case we earn in expectation at least $\scVal(S_1^{(2)})$.  We have proven that $s_2\scVal(S^{(2)}_1)\le\scSrev(S_{-1})$, and by the induction hypothesis $\scRev(S_{-1})\le m\cdot\scSrev(S_{-1})$, so $s_2\scRev(S^{(2)})\le(m+1)\scSrev(S_{-1})$.

Putting everything together, we have $\scRev(S)\le(m+1)(\scSrev(S_1)+\scSrev(S_{-1}))=(m+1)\scSrev(S)$, completing the induction and the proof of the lemma.
\Halmos\endproof

Using these lemmas, we decompose the revenue of the distribution $D$ as follows:
\begin{eqnarray*}
\scRev(D) & \le & \sum_{A\subseteq[n]}p_A\scRev(D_A) \\
& \le & \sum_{A\subseteq[n]}p_A\big(\scVal(D_A^C)+\scRev(D_A^T)\big) \\
& \le & \sum_{A\subseteq[n]}p_A\scVal(D_\emptyset^C)+\sum_{A\subseteq[n]}p_A\scRev(D_A^T) \\
& = & \scVal(D_\emptyset^C)+\sum_{A\subseteq[n]}p_A\scRev(D_A^T) \label{bound_on_rev}
\end{eqnarray*}
where the first inequality is Subdomain Stitching, the second inequality is Marginal Mechanism, the third inequality is immediate from the definition of $D_A^C$, and the equality is a consequence of $\sum_{A\subseteq[n]}p_A=1$.

Now, for all $A\subseteq[n]$ such that $p_A>0$, Lemma~\ref{number_of_items} tells us that $\scRev(D_A^T)\le|A|\scSrev(D_A^T)=|A|\sum_{i\in A}\scSrev(D_i^T)$.  Lemma~\ref{subdomain_reverse} tells us that $\scSrev(D_i^T)\le\frac{r_i}{p_i}$, where $p_i\neq0$ since $p_A>0$, so
\begin{eqnarray*}
\sum_{A\subseteq[n]}p_A\scRev(D_A^T) & \le & \sum_{A\subseteq[n]}p_A|A|\sum_{i\in A}\frac{r_i}{p_i} \\
& = & \sum_{i=1}^nr_i\sum_{A\ni i}|A|\frac{p_A}{p_i} \\
\end{eqnarray*}
$\sum_{A\ni i}|A|\frac{p_A}{p_i}$ is the expected number of items in their tails conditioned on item $i$ being in its tail, so it is equal to $1+\sum_{j\neq i}p_j$.  Thus
\begin{eqnarray*}
\sum_{A\subseteq[n]}p_A\scRev(D_A^T) & \le & \sum_{i=1}^nr_i\Big(1+\sum_{j\neq i}p_j\Big) \\
& = & 1+\sum_{j=1}^np_j\sum_{i\neq j}r_i \\
& = & 1+\sum_{j=1}^np_j(1-r_j)
\end{eqnarray*}

We will use $\tau$ to denote the quantity $\sum_{i=1}^np_i(1-r_i)$.  It is immediate that $\tau\le\sum_{i=1}^np_i\le1$, but we can get a stronger bound for the welfare of the core if we don't immediately apply the inequality $\tau\le1$.  We have
\begin{equation}
\label{tau_bound_on_rev}
\scRev(D)\le\scVal(D_\emptyset^C)+1+\tau
\end{equation}

Before we proceed, one final lemma we will need later is:
\begin{lemma}
\label{variance_lemma}
Let $Y$ be a random variable distributed over $[0,1]$ and suppose $y(1-F(y))$ is upper bounded by some value $v\in[0,1]$.  Then $\mathrm{Var}(Y)\le2v$.
\end{lemma}

\proof{Proof of Lemma~\ref{variance_lemma}.}
\begin{eqnarray*}
\mathrm{Var}(Y) & = & \bE[Y^2]-\bE[Y]^2 \\
& \le & \bE[Y^2] \\
& = & \int_0^1\bP[Y^2\ge y]dy \\
& = & \int_0^1\bP[Y\ge\sqrt{y}]dy \\
& \le & \int_0^1\frac{v}{\sqrt{y}}dy \\
& = & 2v
\end{eqnarray*}
where the second inequality uses the fact that the Myerson revenue for $Y$ is upper bounded by $v$.
\Halmos\endproof

\subsection{A Tighter Bound for the Welfare of the Core}

The main observation behind our improvement is that for $\tau$ to be large (and the above bound to be weak), the tail probabilities must be large.  However, we will choose the price of the grand bundle, $P_t$, to be at most $2$, so that whenever $2$ or more valuations lie in their tails, the customer is guaranteed to want to buy the bundle (and dispose of items for which her valuation is negative).  Thus
\begin{eqnarray}
\bP[\scriptstyle\sum x_i^+<P_t\displaystyle] & = & p_\emptyset\cdot\bP_{x\sim D_\emptyset}[\scriptstyle\sum x_i^+<P_t\displaystyle]+\sum_{|A|=1}p_A\cdot\bP_{x\sim D_A}[\scriptstyle\sum x_i^+<P_t\displaystyle]+\sum_{|A|\ge2}p_A\cdot(0) \nonumber\\
& \le & \Big(p_\emptyset+\sum_{|A|=1}p_A\Big)\cdot\bP_{x\sim D_\emptyset^C}[\scriptstyle\sum x_i^+<P_t\displaystyle] \nonumber\\
& = & \Big(\prod_{i=1}^n(1-p_i)+\sum_{i=1}^np_i\prod_{j\neq i}(1-p_j)\Big)\cdot\bP_{x\sim D_\emptyset^C}[\scriptstyle\sum x_i^+<P_t\displaystyle] \label{tighter_bound_welfare_core}
\end{eqnarray}
where the inequality comes from the fact that the probability of $\sum x_i^+$ being less than the bundle price is greater conditioned on no items being in the tail, than conditioned on some item being in the tail.  We used independence to compute the probabilities in the final expression above, which we will bound in the following way:
\begin{lemma}
\label{key_inequality}
Let $p_1,\ldots,p_n$, $r_1,\ldots,r_n$ be real numbers satisfying $0\le p_i\le r_i$ and $\sum_{i=1}^nr_i=1$.  Let $\tau=\sum_{i=1}^np_i(1-r_i)$.  Then
\begin{equation*}
\prod_{i=1}^n(1-p_i)+\sum_{i=1}^np_i\prod_{j\neq i}(1-p_j)\le\frac{\frac{5}{4}+\tau}{e^\tau}
\end{equation*}
\end{lemma}

This is the key inequality that enables our improved ratio and its proof requires new analysis.  Note that we do indeed have the condition $p_i\le r_i$ in our case, since by Lemma~\ref{subdomain_reverse} $r_i\ge p_i\scRev(D_i^T)$, and $\scRev(D_i^T)$ must be at least $1$ when $D_i^T$ is distributed over $(1,\infty)$.

\proof{Proof of Lemma~\ref{key_inequality}.}
We will first prove
\begin{equation}
\label{three_quarters}
\frac{3}{4}\cdot\prod_{i=1}^n(1-p_i)+\sum_{i=1}^np_i\prod_{j\neq i}(1-p_j)\le\frac{1+\tau}{e^\tau}
\end{equation}
Assume that $p_i<1$ for all $i\in[n]$; the lemma is trivially true otherwise because we would have $\mathrm{LHS}=1$ and $\tau=0$. Since $\tau=\sum_{i=1}^np_i(1-r_i)$ and $1-x\le e^{-x}$, it suffices to prove
\begin{equation*}
\frac{3}{4}\cdot\prod_{i=1}^n(1-p_i)+\sum_{i=1}^np_i\prod_{j\neq i}(1-p_j)\le\Big(1+\sum_{i=1}^np_i(1-r_i)\Big)\prod_{i=1}^n(1-p_i(1-r_i))
\end{equation*}
which is equivalent to
\begin{equation*}
\frac{3}{4}+\sum_{i=1}^n\frac{p_i}{1-p_i}\le\Big(1+\sum_{i=1}^n(p_i-p_ir_i)\Big)\prod_{i=1}^n(1+\frac{p_ir_i}{1-p_i})
\end{equation*}
Observe that the RHS is at least
\begin{eqnarray*}
& & \Big(1+\sum_{i=1}^n(p_i-p_ir_i)\Big)\Big(1+\sum_{i=1}^n\frac{p_ir_i}{1-p_i}\Big) \\
& = & 1+\sum_{i=1}^n\frac{(p_i-p_ir_i)(1-p_i)+p_ir_i}{1-p_i}+\Big(\sum_{i=1}^np_i(1-r_i)\Big)\Big(\sum_{i=1}^n\frac{p_ir_i}{1-p_i}\Big) \\
& = & 1+\sum_{i=1}^n\frac{p_i}{1-p_i}-\sum_{i=1}^n\frac{p_i^2(1-r_i)}{1-p_i}+\Big(\sum_{i=1}^np_i(1-r_i)\Big)\Big(\sum_{i=1}^n\frac{p_ir_i}{1-p_i}\Big) \\
& = & 1+\sum_{i=1}^n\frac{p_i}{1-p_i}-\sum_{i=1}^n\frac{p_i^2(1-r_i)^2}{1-p_i}+\sum_{i\neq j}p_i(1-r_i)\cdot\frac{p_jr_j}{1-p_j} \\
\end{eqnarray*}
so it remains to prove
\begin{equation*}
\sum_{i=1}^n\frac{p_i^2(1-r_i)^2}{1-p_i}-\sum_{i\neq j}p_i(1-r_i)\cdot\frac{p_jr_j}{1-p_j}\le\frac{1}{4}
\end{equation*}
But $p_i\le r_i$ for all $i\in[n]$, so the LHS is at most $\sum_{i=1}^np_i^2(1-p_i)$, which can be seen to be at most $\frac{1}{4}$, since $p_i(1-p_i)$ is always at most $\frac{1}{4}$ and $\sum_{i=1}^np_i\le1$.

Also, since $\tau\le\sum_{i=1}^np_i$, $e^{-\tau}\ge\exp(-\sum_{i=1}^np_i)\ge\prod_{i=1}^n(1-p_i)$.  Multiplying by $\frac{1}{4}$ and adding to (\ref{three_quarters}), we complete the proof of the lemma.
\Halmos\endproof

\subsection{Applying Concentration Inequalities}

To bound $\bP_{x\sim D_\emptyset^C}[\sum x_i^+<P_t]$, we want to show that $\sum x_i^+$ concentrates around its mean, where valuation $x_i$ is drawn from its conditional core distribution $D_i^C$ for all $i\in[n]$.  Note that $y(1-F_{x_i}(y))$ is bounded above by $r_i$ for all $y\in[0,1]$; otherwise $\scSrev(D_i^C)>r_i\implies\scSrev(D_i)>r_i$ which is a contradiction.  Hence $y(1-F_{x_i^+}(y))$ is also bounded above by $r_i$ and we can invoke Lemma~\ref{variance_lemma} to get $\mathrm{Var}_{x_i\sim D_i^C}(x_i^+)\le2r_i$ for all $i\in[n]$.  By independence, $\mathrm{Var}_{x\sim D_\emptyset^C}(\sum x_i^+)=\sum_{i=1}^n\mathrm{Var}_{x\sim D_\emptyset^C}(x_i^+)\le\sum_{i=1}^n2r_i=2$ and we have successfully bounded the variance of the quantity we are interested in.


We again use Cantelli's inequality (see Lemma~\ref{cantelli}) for bounding the one-sided tail.
Note that $\bE_{x\sim D_\emptyset^C}[\sum_{i=1}^nx_i^+]=\scVal(D_\emptyset^C)$ by definition.  Also, it will be convenient to write the bundle price as $P_t=\alpha\cdot\scVal(D_\emptyset^C)$, for some $\alpha\in[0,1]$ (we would never want $\alpha>1$ since then the price would be greater than the mean and it would be impossible to use Cantelli's inequality).  Then Cantelli's inequality tells us that
\begin{eqnarray*}
\bP_{x\sim D_\emptyset^C}[\scriptstyle\sum x_i^+<P_t\displaystyle] & = & \bP_{x\sim D_\emptyset^C}\Big[\sum_{i=1}^nx_i^+-\scVal(D_\emptyset^C)<-(1-\alpha)\scVal(D_\emptyset^C)\Big] \\
& \le & \frac{\mathrm{Var}_{x\sim D_\emptyset^C}(\sum x_i^+)}{\mathrm{Var}_{x\sim D_\emptyset^C}(\sum x_i^+)+(1-\alpha)^2\scVal(D_\emptyset^C)^2} \\
& \le & \frac{2}{2+(1-\alpha)^2\scVal(D_\emptyset^C)^2} \\
\end{eqnarray*}
where he second inequality comes from our variance bound above.  So long as we choose $P_t\le2$, we can use (\ref{tighter_bound_welfare_core}), and combined with Lemma~\ref{key_inequality} we get
\begin{equation*}
\bP[\scriptstyle\sum x_i^+<P_t\displaystyle]\le\min\big\{\frac{1.25+\tau}{e^\tau},1\big\}\cdot\frac{2}{2+(1-\alpha)^2\scVal(D_\emptyset^C)^2}
\end{equation*}
and hence the expected revenue from selling the grand bundle at price $\alpha\cdot\scVal(D_\emptyset^C)$ is at least
\begin{equation*}
\alpha\cdot\scVal(D_\emptyset^C)\cdot\Big(1-\min\big\{\frac{1.25+\tau}{e^\tau},1\big\}\cdot\frac{2}{2+(1-\alpha)^2\scVal(D_\emptyset^C)^2}\Big)
\end{equation*}
Recall from (\ref{tau_bound_on_rev}) that $\scRev(D)\le\scVal(D_\emptyset^C)+1+\tau$.  While $\tau$ could take on any value in $[0,1]$, we can choose the price of the bundle based on $\tau$ and $\scVal(D_\emptyset^C)$ by adjusting $\alpha\in[0,1]$.

\noindent\textbf{Case 1.} If $\scVal(D_\emptyset^C)\le3.2$, then $\scRev(D)\le3.2+1+1=5.2\cdot\scSrev(D)$ is immediate and we can just sell the items individually.

\noindent\textbf{Case 2.} If $3.2<\scVal(D_\emptyset^C)\le4$, then we will choose $\alpha=\frac{1}{2}$ which guarantees $P_t\le2$.  Thus
\begin{equation*}
\scBdcrev(D)\ge\scVal(D_\emptyset^C)\cdot\frac{1}{2}\Big(1-\min\big\{\frac{1.25+\tau}{e^\tau},1\big\}\cdot\frac{2}{2+(1-\frac{1}{2})^2(3.2)^2}\Big)
\end{equation*}
It can be shown with calculus that:
\begin{proposition}
For all $\tau\in[0,1]$, $2\Big(1-\min\left\{\frac{1.25+\tau}{e^\tau},1\right\}\cdot\frac{2}{2+(1-\frac{1}{2})^2(3.2)^2}\Big)^{-1}+(1+\tau)<5.2$, with the maximum of $\approx5.1952$ occuring at the unique positive $\tau$ satisfying $\frac{1.25+\tau}{e^\tau}=1$.
\end{proposition}
Hence $\scVal(D_\emptyset^C)\le(4.2-\tau)\scBdcrev(D)$.  Substituting into (\ref{tau_bound_on_rev}), we get
\begin{eqnarray*}
\scRev(D) & \le & (4.2-\tau)\scBdcrev(D)+(1+\tau)\scSrev(D) \\
& \le & 5.2\cdot\max\{\scSrev(D),\scBdcrev(D)\}
\end{eqnarray*}
as desired.

\noindent\textbf{Case 3.} If $4<\scVal(D_\emptyset^C)$, then we will still choose $\alpha=\frac{1}{2}$.  We no longer have $P_t\le2$, so we have to use the weaker bound $\bP_{x\sim D}[\sum x_i^+<P_t]\le\bP_{x\sim D_\emptyset^C}[\sum x_i^+<P_t]$.  However, applying Cantelli yields
\begin{equation*}
\bP_{x\sim D_\emptyset^C}[\scriptstyle\sum x_i^+<P_t\displaystyle]\le\frac{2}{2+(1-\frac{1}{2})^2(4)^2}=\frac{1}{3}
\end{equation*}
so $\scBdcrev(D)\ge\scVal(D_\emptyset^C)\cdot\frac{1}{2}(1-\frac{1}{3})$.  We get $\scRev(D)\le3\cdot\scBdcrev(D)+(1+\tau)\scSrev(D)<5.2\cdot\max\{\scSrev(D),\scBdcrev(D)\}$.

This concludes the proof of Theorem~\ref{mr2}.  In the next subsection we prove our upper bound Theorem~\ref{thm_upper_bound}.

\subsection{Proof of Theorem~\ref{thm_upper_bound} from Section~\ref{sec_upper_bound}}

It is immediate that the optimal revenue from PC is $2\rho$, attained by selling individual items at any price in $[1,2]$.  Next, we would like to argue that the optimal revenue from PB is also $2\rho$.  If we offer the bundle at $2$, it is guaranteed to get bought if either valuation realizes to $2$ or both valuations realize to a positive number, and won't get bought otherwise.  Therefore the revenue is $2(\rho^2+2(1-\rho)\frac{\rho}{2})=2\rho$.

We can do equally well by offering the bundle at $3$, and any other price is inferior.
\begin{lemma}
\label{upper_bound}
The optimal revenue from PB is $2\rho$, attained by setting a bundle price of $2$ or $3$.
\end{lemma}

\proof{Proof of Lemma~\ref{upper_bound}.}
Let $z$ denote the price of the bundle.  We will systematically analyze all the cases over $1\le z\le4$ and show that the maximum revenue of $2\rho$ is attained at $z=2$ and $z=3$.

\noindent\textbf{Case 1.} Suppose $1\le z\le2$.  Let us condition on the realization $y$ of the first valuation.  If $y=0$, then we get a sale with probability $\frac{\rho}{z}$.  If $y\in[1,z)$, then we get a sale so long as the second valuation realizes to a positive number, which occurs with probability $1-\rho$.  If $y\ge z$, then the first valuation alone is enough to guarantee a bundle sale.  The expected revenue is
\begin{equation*}
z\left((1-\rho)\frac{\rho}{z}+(\rho-\frac{\rho}{z})\rho+\frac{\rho}{z}\right)=2\rho+(z-2)\rho^2
\end{equation*}
which is clearly maximized at $z=2$, in which case the revenue is $2\rho$.

\noindent\textbf{Case 2.} Suppose $2<z\le3$.  Let us condition on the realization $y$ of the first valuation.  If $y=0$, then we have no chance of selling the bundle.  If $y\in[1,z-1]$, then we get a sale when the other valuation is at least $z-y$.  Since $z-y\in[1,2]$, the probability of this occurring is $\frac{\rho}{z-y}$.  If $y\ge z-1$, then we get a sale so long as the other valuation realizes to a positive number, which occurs with probability $\rho$.  The total probability of getting a sale is
\begin{equation*}
\int_1^{z-1}\frac{\rho}{y^2}\frac{\rho}{z-y}dy+\frac{\rho}{z-1}\rho
\end{equation*}
where the PDF of $Y$ satisfies $f(y)=\frac{\rho}{y^2}$ over $[1,2)$.  Using partial fractions, the antiderivative of $\frac{1}{y^2(z-y)}$ can be computed to be
\begin{equation*}
\frac{1}{z}\left(\frac{\ln y-\ln(z-y)}{z}-\frac{1}{y}\right)
\end{equation*}
as demonstrated in the proof of \cite[lem.~6]{hart2017approximate}.  Therefore, the definite integral evaluates to
\begin{equation*}
\rho^2\left(\frac{2\ln(z-1)}{z^2}+\frac{2}{z}-\frac{1}{z-1}\right)
\end{equation*}
and the expected revenue is
\begin{equation*}
z\rho^2\left(\frac{2\ln(z-1)}{z^2}+\frac{2}{z}-\frac{1}{z-1}+\frac{1}{z-1}\right)=2\rho^2\left(\frac{\ln(z-1)}{z}+1\right)
\end{equation*}
However, $\frac{\ln(z-1)}{z}$ is a strictly increasing function on $(2,3]$, so this expression is uniquely maximized at $z=3$ where it equals $2\rho^2(\frac{\ln2}{3}+1)=2\rho$.

\noindent\textbf{Case 3.} Suppose $3\le z\le4$.  Let us condition on the realization $y$ of the first valuation.  If $y<z-2$, then we have no chance of selling the bundle.  Otherwise, the probability of getting a sale is $\frac{\rho}{z-y}$, since $z-y\in[1,2]$.  The total probability of getting a sale is
\begin{equation*}
\int_{z-2}^2\frac{\rho}{y^2}\frac{\rho}{z-y}dy+\frac{\rho}{2}\frac{\rho}{z-2}
\end{equation*}
and the integral evaluates to
\begin{equation*}
\rho^2\left(\frac{2\ln2-2\ln(z-2)}{z^2}+\frac{1}{z(z-2)}-\frac{1}{2z}\right)
\end{equation*}
Therefore, the expected revenue is
\begin{equation*}
z\rho^2\left(\frac{2\ln2-2\ln(z-2)}{z^2}+\frac{1}{z(z-2)}-\frac{1}{2z}+\frac{1}{2(z-2)}\right)=2\rho^2\left(\frac{\ln2-\ln(z-2)}{z}+\frac{1}{z-2}\right)
\end{equation*}
$\frac{\ln2-\ln(z-2)}{z}+\frac{1}{z-2}$ is a strictly decreasing function on $[3,4]$, so this expression is uniquely maximized at $z=3$.
\Halmos\endproof

Now, consider the strategy of offering either item for $2$ or the bundle for the discounted price of $3$.  Note that if buying the bundle is non-negative utility for the customer, then buying either individual item cannot be higher utility, since the price savings is one and the value of the item lost is at least one (recall that the firm gets to break ties in a way that favors itself).  Hence there is no cannibalization of bundle sales from individual sales and we earn revenue at least $2\rho$.  However, when exactly one valuation realizes to a positive number (in which case we have no chance of selling the bundle), we still have a $\frac{1}{2}$ conditional probability of selling that individual item.  Hence the revenue from Mixed Bundling is $2\rho+2(2(1-\rho)\frac{\rho}{2})=2\rho(2-\rho)$.

The relative gain over both the PC revenue and the PB revenue is $2-\rho=\frac{3+2\ln2}{3+\ln2}$, completing the proof of Theorem~\ref{thm_upper_bound}.

\section{Example where BSP Performs Poorly}\label{video_game_example}

Consider a firm that is bundling a higher-profit-margin, lower-valuation good with a low-profit-margin, high-valuation good.  This is a common occurrence, for example when video games are bundled with a console, which we will hereinafter refer to as item~1 and item~2, respectively.  Item~1 costs zero to produce and has a valuation uniform on [0,1]; item~2 costs 4.5 to produce and has a valuation uniform on [0,5] and independent from item~1.  Most of the welfare comes from the lower-valuation item: the expected welfare for item~1 and item~2 are 0.5 and 0.025, respectively.

The optimal deterministic profit is $\approx0.265$, attained by offering item~1 at 0.51, item~2 at 4.83, and the bundle at the discounted price of 5.13.

The optimal BSP pricing charges 4.83 for a single item and 5.03 for both items, earning only 19\% of the deterministic optimum.  This example highlights the issue with BSP: it cannot afford to charge a low price for a single item if any item has a high production cost.  However, most of the potential profit could be coming from offering lower-valuation items at low prices!  \cite{CLS08} bypass such examples in their numerical experiments, assuming that all items have a low cost compared to its mean valuation.

PBDC offers item~1 at 0.51, item~2 at 5.01, and the bundle at 5.01---which is the right idea and earns 99.1\% of the deterministic optimum.  Interestingly, even the analytical solution provided by \cite{Bha13}, which computes the optimal deterministic pricing when there are two independent uniform distributions and costs, is less effective than PBDC on this example.  The solution from \cite{Bha13} only attains $97.5\%$ of the deterministic optimum for this example, because it requires a bit of linear approximation.

Optimal bundling is an intricate problem even in the case of two independent uniform distributions, so a pricing heuristic as robust as PBDC is invaluable.  In fact, for this example PBDC recommends \emph{Partial Mixed Bundling}, which is a Mixed Bundling scheme where one of the items, in this case item 2 (the high-cost low-welfare item), is never sold individually.  This matches the intuition that the seller should add item 1 (the low-cost high-welfare item) to item 2 in order to increase the total amount customer is willing to pay (see Proposition~1 in \cite{Bha13}).  BSP, on the other hand, does not perform well: it recommends a Partial Mixed Bundling scheme where item 1 is never sold individually.


\end{document}